\documentstyle[prd,aps,preprint,tighten,floats,epsf]{revtex} 

\input{epsf}

\begin{document}

\reversemarginpar

\title{Interpolating between the Bose-Einstein and\\ 
the Fermi-Dirac distributions in odd dimensions}
\author{L.~Sriramkumar~\thanks{E-mail:~sriram@mri.ernet.in}}
\address{Harish-Chandra Research Institute, Chhatnag Road\\ 
Jhunsi, Allahabad 211 019, India.}

\maketitle

\begin{abstract}
We consider the response of a uniformly accelerated monopole 
detector that is coupled to a {\it superposition}\/ of an odd 
and an even power of a quantized, massless scalar field in flat 
spacetime in arbitrary dimensions. 
We show that, when the field is assumed to be in the Minkowski 
vacuum, the response of the detector is characterized by a 
Bose-Einstein factor in even spacetime dimensions, whereas a 
Bose-Einstein {\it as well as}\/ a Fermi-Dirac factor appear 
in the detector response when the dimension of spacetime is odd.
Moreover, we find that, it is possible to {\it interpolate}\/ 
between the Bose-Einstein and the Fermi-Dirac distributions in odd 
spacetime dimensions by suitably adjusting the relative strengths 
of the detector's coupling to the odd and the even powers of the 
scalar field.   
We point out that the response of the detector is always thermal
and we, finally, close by stressing the {\it apparent}\/ nature of 
the appearance of the Fermi-Dirac factor in the detector response.
\end{abstract}

\newpage

\section{Introduction}
It is well-known that the response of a uniformly accelerated 
Unruh-DeWitt detector that is coupled to a quantized, massless 
scalar field is characterized by a Planckian distribution when the 
field is assumed to be in the Minkowski vacuum\cite{unruh76,dewitt79}.
However, what does not seem to be so commonly known is the fact 
that this result is true only in even spacetime dimensions, and, 
in odd spacetime dimensions, a Fermi-Dirac factor (rather than a 
Bose-Einstein factor) appears in the response of the accelerated 
Unruh-DeWitt detector (for the original results, see Refs.
\cite{stephens85,takagi84,takagi85a,stephens86,ooguri86,unruh86,takagi86}; 
for relatively recent discussions, see Refs.~\cite{anglin93,terashima99}). 

The Unruh-DeWitt detector is a monopole detector that is coupled 
{\it linearly}\/ to the quantum scalar field\cite{unruh76,dewitt79}. 
With a variety of motivations in mind, there has been a prevailing 
interest in literature in studying the response of detectors that 
are coupled {\it non-linearly}\/ to the quantum 
field\cite{hinton83,hinton84,paddytp87,suzuki97,sriram99}. 
In a recent Letter\cite{sriram02}, we had considered the response 
of a uniformly accelerated monopole detector that is coupled to 
an arbitrary (but, positive) integral power of a massless, quantum 
scalar field in $(D+1)$-dimensional flat spacetime.
We had found that, when the detector is coupled to an even power of 
the scalar field, a Bose-Einstein factor arises in the response of 
the detector (in the Minkowski vacuum) in {\it all}\/ spacetime 
dimensions, whereas, a Fermi-Dirac factor appears in the detector 
response {\it only}\/ when {\it both}\/ the spacetime dimension 
[viz.~$(D+1)$] and the index of non-linearity of the coupling are 
odd.

In this note, we shall consider the response of an accelerated 
monopole detector that is coupled to a {\it superposition}\/ of 
an odd and an even power of the massless, quantum scalar field.
Though the response of such a detector in the Minkowski vacuum is 
characterized by the Bose-Einstein factor in even spacetime dimensions, 
interestingly, we find that, in odd spacetime dimensions, the response 
of the detector contains an admixture of the Bose-Einstein and the 
Fermi-Dirac factors.
Also, as we shall see, it is possible to interpolate between the 
Bose-Einstein and the Fermi-Dirac factors in odd spacetime
dimensions by suitably modulating the relative strengths of the 
detector's coupling to the odd and the even powers of the quantum 
scalar field. 
In what follows, we shall set $\hbar=c=k_{\rm B}=1$ and, for convenience 
in notation, denote the trajectory $x^{\mu}(\tau)$ of the detector as 
${\tilde x}(\tau)$ with $\tau$ being the proper time in the frame 
of the detector.

\section{``Inverted statistics'' for odd couplings}
Let us begin by reviewing our earlier result for a monopole detector 
that is coupled to the $n$th power (with $n$ being a positive integer) 
of a real scalar field~$\Phi$ through the following interaction 
Lagrangian\cite{suzuki97,sriram02}:
\begin{equation}
{\cal L}_{\rm NL}
= {\bar c}\, m(\tau)\; \Phi^n\left[{\tilde x}(\tau)\right],
\label{eqn:nlint}
\end{equation}
where ${\bar c}$ is a small coupling constant and $m(\tau)$ is 
the detector's monopole moment.
Consider a situation wherein the quantum field ${\hat \Phi}$ is 
initially in the vacuum state $\left\vert 0 \right\rangle$ 
and the detector is in its ground state~$\left\vert E_0 
\right\rangle$ corresponding to an energy eigen value~$E_0$.
Then, up to the first order in perturbation theory, the 
amplitude of transition of the detector to an excited 
state~$\left\vert E \right\rangle$, corresponding to an 
energy eigen value~$E\, \left(>E_0\right)$, is described 
by the integral~\cite{suzuki97,sriram02} 
\begin{equation}
{\cal A}_{n}({\cal E}) 
=\left(i{\bar c} {\cal M}\right)\, 
\int\limits_{-\infty}^{\infty} d\tau\, e^{i {\cal E}\tau}\, 
\left\langle\Psi\right\vert
:{\hat \Phi}^n[{\tilde x}(\tau)]:\left\vert 0\right\rangle,
\label{eqn:nonldetamp}
\end{equation}
where ${\cal M}\equiv \left\langle E \right\vert {\hat m}(0)
\left\vert E_{0} \right\rangle$, ${\cal E}=\left(E-E_0\right)>0$, 
$\left\vert \Psi \right\rangle$ is the state of the quantum scalar 
field after its interaction with the detector and the colons denote
normal ordering with respect to the Minkowski vacuum.
(The normal ordering procedure is required to overcome the divergences 
that would otherwise arise for $n>1$. For a detailed discussion on this 
point, see Refs.~\cite{suzuki97,sriram02}.)
The transition probability of the detector to all possible final 
states $\left\vert \Psi\right\rangle$ of the quantum field is then
given by
\begin{equation}
{\cal P}_{n}({\cal E}) 
=\sum_{\vert\Psi\rangle}{\vert {\cal A}_{n}({\cal E})\vert}^2
= \int\limits_{-\infty}^\infty d\tau\, 
\int\limits_{-\infty}^\infty d\tau'\, 
e^{-i{\cal E}(\tau-\tau')}\, 
G^{(n)}\left[{\tilde x}(\tau), {\tilde x}(\tau')\right],
\label{eqn:nldetprob}
\end{equation}
where  we have dropped an (irrelevant) overall factor of 
$\left(\vert {\bar c} \vert\vert {\cal M}\vert\right)^2$ 
and $G^{(n)}\left[{\tilde x}(\tau), {\tilde x}(\tau')\right]$ 
is a $(2n)$-point function defined as
\begin{equation}
G^{(n)}\left[{\tilde x}(\tau), {\tilde x}(\tau')\right]
=\left\langle 0 \right\vert 
:{\hat \Phi}^n\left[{\tilde x}(\tau)\right]:\,
:{\hat \Phi}^n\left[{\tilde x}(\tau')\right]:
\left\vert 0 \right\rangle.\label{eqn:2nptfn}
\end{equation}
For trajectories wherein the $(2n)$-point function 
$G^{(n)}\left(\tau, \tau'\right) (\equiv G^{(n)}
\left[{\tilde x}(\tau), {\tilde x}(\tau')\right])$ is invariant under 
translations in the proper time in the frame of the detector, as in the 
case of the Unruh-DeWitt detector, we can define a transition probability 
rate for the non-linearly coupled detector as follows:
\begin{equation}
{\cal R}_{n}({\cal E}) 
= \int\limits_{-\infty}^\infty d{\bar \tau}\;
e^{-i{\cal E}{\bar \tau}}\; G^{(n)}({\bar \tau}),
\label{eqn:nldetrate}
\end{equation} 
where ${\bar \tau}=(\tau -\tau')$.

If we assume that the quantum field is in the Minkowski vacuum, 
then, using Wick's theorem, it is easy to show that, 
the $(2n)$-point function $G^{(n)}\left({\tilde x}, {\tilde x'}
\right)$ above simplifies to~\cite{sriram02}
\begin{equation}
G^{(n)}_{\rm M}\left({\tilde x}, {\tilde x'}\right)
= \left(n!\right)\, \left[G^{+}_{\rm M}\left({\tilde x}, 
{\tilde x'}\right)\right]^n,\label{eqn:2nptfnmv}
\end{equation}
where $G^{+}_{\rm M}\left({\tilde x}, {\tilde x'}\right)$ 
denotes the Wightman function in the Minkowski vacuum.
Along the trajectory of a detector that is accelerating uniformly 
with a proper acceleration~$g$ in a particular direction, the 
Wightman function for a massless scalar field in the Minkowski 
vacuum in $(D+1)$ spacetime dimensions (for $(D+1)\ge 3$) is 
given by\cite{takagi85a,unruh86,takagi86}
\begin{equation}
G_{\rm M}^{+}\left({\bar \tau}\right)
=\left[{\cal C}_{D}\; (g/2i)^{(D-1)}\right]\;
\biggl({\rm sinh}\left[\left(g{\bar \tau}/2\right)
-i\epsilon\right]\biggl)^{-(D-1)},
\label{eqn:wgfnmvrind}
\end{equation}
where $\epsilon\to 0^{+}$ and ${\cal C}_{D}
= \left[\Gamma\left[(D-1)/2\right]/\left(4\pi^{(D+1)/2}\right)\right]$ 
with $\Gamma\left[(D-1)/2\right]$ denoting the Gamma function.
Therefore, along the trajectory of the uniformly 
accelerated detector, the $(2n)$-point function in 
the Minkowski vacuum~(\ref{eqn:2nptfnmv}) reduces to
\begin{equation}
G^{(n)}_{\rm M}\left({\bar \tau}\right)
=(n!)\;\left[{\cal C}_{D}^{n}\; (g/2i)^{\alpha}\right]\; 
\biggl({\rm sinh}\left[\left(g{\bar \tau}/2\right)
-i\epsilon\right]\biggl)^{-\alpha},
\label{eqn:2nptfnmvrind}
\end{equation}
where $\alpha=\left[(D-1)n\right]$.
On substituting this $(2n)$-point function in the 
expression~(\ref{eqn:nldetrate}) and carrying out the 
integral, we find that the transition probability 
rate of the uniformly accelerated, non-linearly
coupled detector can be expressed as\cite{sriram02}
\begin{equation}
{\cal R}_{n}({\cal E}) 
={\cal B}(n,D)\;\;
\left\{
\begin{array}{l}
\left(g^{\alpha}/{\cal E}\right)\;
{\underbrace{\left[\exp(2\pi{\cal E}/g)-1\right]^{-1}}}\;
\prod\limits_{l=0}^{(\alpha-2)/2}
\left[l^2+({\cal E}/g)^2\right]\\
\qquad\quad\,\mbox{Bose-Einstein factor}\qquad
\qquad\qquad\quad\;\mbox{when $\alpha$ is even}\\
g^{(\alpha-1)}\;\;
{\underbrace{\left[\exp(2\pi{\cal E}/g)+1\right]^{-1}}}\;
\prod\limits_{l=0}^{(\alpha-3)/2}
\left[((2l+1)/2)^2+({\cal E}/g)^2\right]\\
\qquad\quad\;\mbox{Fermi-Dirac factor}\qquad
\qquad\qquad\qquad\mbox{when $\alpha$ is odd,}
\end{array}\right.
\label{eqn:nldetraterind}
\end{equation}
where the quantity ${\cal B}(n,D)$ is given by
\begin{equation}
{\cal B}(n,D)=(2\pi)\,(n!)\; 
\left[{\cal C}_{D}^{n}/\Gamma(\alpha)\right].
\end{equation} 
Since, for even $(D+1)$, $\alpha$ is even for all $n$, a Bose-Einstein 
factor will always arise in the response of the uniformly accelerated 
detector in even-dimensional flat spacetimes.
On the other hand, when $(D+1)$ is odd, clearly, $\alpha$ will be 
odd or even depending on whether $n$ is odd or even.
As a result, in odd-dimensional flat spacetimes, a Bose-Einstein 
factor will arise in the detector response only when~$n$ is even, 
but, as in the case of the Unruh-DeWitt detector, a Fermi-Dirac
factor will appear when~$n$ is odd.
[Note that the temperature associated with the Bose-Einstein and the 
Fermi-Dirac factors is the standard Unruh temperature, viz.~$(g/2\pi)$.]
Also, the response of the detector will be characterized completely by 
either a Bose-Einstein or a Fermi-Dirac distribution only in cases such 
that $\alpha<3$ and, in situations wherein $\alpha\ge 3$, the detector 
response will contain, in addition to a Bose-Einstein or a Fermi-Dirac 
factor, a term which is polynomial in $({\cal E}/g)$.

\section{``Mixing statistics'' with a superposition of\\ 
odd and even couplings}
Now, consider a detector that interacts with the scalar field $\Phi$ 
through the following Lagrangian:
\begin{equation}
{\cal L}_{\rm SP}
= m(\tau)\; \biggl({\bar c}_{\rm o}\,  
\Phi^{n_{\rm o}}\left[{\tilde x}(\tau)\right]
+ {\bar c}_{\rm e}\, \Phi^{n_{\rm e}}\left[{\tilde x}(\tau)\right]
\biggl),
\end{equation}
where ${\bar c}_{\rm o}$ and ${\bar c}_{\rm e}$ denote two (small) 
coupling constants and $n_{\rm o}$ and $n_{\rm e}$ are positive integers. 
The transition amplitude of such a detector [under the same conditions 
as in the case of ${\cal A}_{\rm n}({\cal E})$] up to the first order 
in the coupling constants ${\bar c}_{\rm o}$ and ${\bar c}_{\rm e}$ will 
be given by
\begin{equation}
{\cal A}_{\rm SP}({\cal E}) = 
\left(i{\cal M}\right)\,
\int\limits_{-\infty}^{\infty} d\tau\, e^{i {\cal E}\tau}\,
\biggl({\bar c}_{\rm o}\, \left\langle\Psi\right\vert\,
:{\hat \Phi}^{n_{\rm o}}[{\tilde x}(\tau)]:\,
\left\vert 0\right\rangle
+{\bar c}_{\rm e}\, \left\langle\Psi\right\vert\,: 
{\hat \Phi}^{n_{\rm e}}[{\tilde x}(\tau)]:\,
\left\vert 0\right\rangle\biggl),
\end{equation}
where, as before, we have normal-ordered the matrix-elements 
in order to avoid the divergences.
If we now assume that $n_{\rm o}$ is odd and $n_{\rm e}$ is even, 
then, being the expectation values of an odd power 
[viz.~$\left(n_{\rm o}+n_{\rm e}\right)$] of the quantum field, 
the cross terms in the corresponding transition probability vanish.
As a result, we obtain that
\begin{equation}
{\cal P}_{\rm SP}({\cal E})
= \int\limits_{-\infty}^\infty d\tau\, 
\int\limits_{-\infty}^\infty d\tau'\, 
e^{-i{\cal E}(\tau-\tau')}\, 
\biggl(G^{(n_{\rm e})}\left[{\tilde x}(\tau), {\tilde x}(\tau')\right]
+ r^2\; 
G^{(n_{\rm o})}\left[{\tilde x}(\tau), {\tilde x}(\tau')\right]\biggl),
\end{equation}
where $G^{(n_{\rm e})}\left[{\tilde x}(\tau), {\tilde x}(\tau')\right]$ 
and $G^{(n_{\rm o})}\left[{\tilde x}(\tau), {\tilde x}(\tau')\right]$ are 
the $(2n)$-point functions [as defined in Eq.~(\ref{eqn:2nptfn})] 
corresponding to $n_{\rm e}$ and $n_{\rm o}$ and the quantity 
$r=\left(\vert {\bar c}_{\rm o}\vert/\vert {\bar c}_{\rm e}\vert\right)$ 
denotes the relative strength of the detector's coupling to the odd power 
of the scalar field with respect to the even power.
[As we had done earlier, in the above expression for ${\cal P}_{\rm SP}
({\cal E})$, we have dropped an unimportant overall factor of 
$\left(\vert {\bar c}_{\rm e}\vert {\vert\cal M}\vert\right)^2$.]
In situations wherein the $(2n)$-point functions are invariant under 
translations in the detector's proper time, the transition probability 
rate of the detector can be expressed as
\begin{equation}
{\cal R}_{\rm SP}({\cal E}) 
= \biggl[{\cal R}_{n_{\rm e}}({\cal E})+
r^2\; {\cal R}_{n_{\rm o}}({\cal E})\biggl],
\end{equation}
where ${\cal R}_{n}({\cal E})$ denotes the transition probability 
rate defined in Eq.~(\ref{eqn:nldetrate}). 

Therefore, for a detector that is in motion along a uniformly 
accelerated trajectory, when the field is assumed to be in the 
Minkowski vacuum, the quantities ${\cal R}_{n_{\rm e}}({\cal E})$ 
and ${\cal R}_{n_{\rm o}}({\cal E})$ in the expression for 
${\cal R}_{\rm SP}({\cal E})$ above will be given by 
${\cal R}_{n}({\cal E})$ in Eq.~(\ref{eqn:nldetraterind}) 
corresponding to the even and the odd integers~$n_{\rm e}$ and 
$n_{\rm o}$, respectively.
Evidently, in such a case, in even spacetime dimensions, both 
${\cal R}_{n_{\rm e}}({\cal E})$ and ${\cal R}_{n_{\rm o}}({\cal E})$ 
will be characterized by a Bose-Einstein factor.
Whereas, in odd spacetime dimensions, ${\cal R}_{\rm SP}({\cal E})$ 
will contain an admixture of the two distributions, with 
${\cal R}_{n_{\rm e}}({\cal E})$ being characterized by a Bose-Einstein 
factor, while ${\cal R}_{n_{\rm o}}({\cal E})$ contains a Fermi-Dirac 
factor. 
Moreover, in odd spacetime dimensions, the detector response 
function ${\cal R}_{\rm SP}({\cal E})$ can be {\it interpolated}\/ 
between the Bose-Einstein and the Fermi-Dirac distributions by 
varying the quantity~$r$ (viz.~the relative strength of the coupling 
constant ${\bar c}_{\rm o}$ with respect to ${\bar c}_{\rm e}$) from 
zero to infinity.

\section{Discussion}
An important point needs to be stressed regarding the appearance 
of the Fermi-Dirac factor in the response of a detector that is 
coupled to a scalar field. 
According to principle of detailed balance, a spectrum 
${\cal R}_{\beta}({\cal E})$ can be considered to be a 
thermal distribution at the inverse temperature $\beta$ 
if the spectrum satisfies the following condition (see, 
for e.g., Refs.~\cite{ooguri86,takagi86}): 
\begin{equation}
{\cal R}_{\beta}({\cal E})
=\left[e^{-\beta {\cal E}}\, {\cal R}_{\beta}(-{\cal E})\right]. 
\end{equation}
It is straightforward to check that this condition is always satisfied 
by the detector response functions ${\cal R}_{n}({\cal E})$ and 
${\cal R}_{\rm SP}({\cal E})$ along the uniformly accelerated trajectory.
Clearly, in spite of the appearance of polynomial terms as well as 
an admixture of the Bose-Einstein and the Fermi-Dirac factors, the 
response of the detectors is indeed thermal.

Actually, the principle of detailed balance is a consequence of the
Kubo-Martin-Schwinger (KMS) condition according to which the Wightman
function of a Bosonic field in thermal equilibrium at the inverse
temperature $\beta$ should be skew-periodic in imaginary proper time 
with a period $\beta$. 
(The Wightman function of a Fermionic field would be skew {\it and}\/ 
anti-periodic in such a situation. For a discussion on this point, 
see, for instance, Refs.~\cite{ooguri86,takagi86}).
It is straightforward to check that, along the uniformly accelerated 
trajectory, the Wightman function in the Minkowski 
vacuum~(\ref{eqn:wgfnmvrind}) indeed satisfies the KMS condition 
(corresponding to the Unruh temperature) as required for a 
{\it Bosonic}\/ field in {\it all}\/ dimensions.
If the Wightman function satisfies the KMS condition of a scalar 
field, then, obviously, all $(2n)$-point functions as well as a 
linear superposition of such functions that are constructed out 
of the Wightman function will also satisfy the same KMS condition.
This immediately suggests that the ``inversion of statistics'',
i.e. the appearance of a Fermi-Dirac factor in the response 
of a detector that is coupled to a scalar field, is an {\it apparent}\/ 
phenomenon---it reflects a curious aspect of these detectors rather 
than point to any fundamental change in statistics in odd spacetime 
dimensions (it is for this reason that we have referred to statistics 
within quotes in the titles of the last two sections). 
Nevertheless, when models with (large and compact) extra dimensions are 
in vogue in literature, the fact that the characteristic response of 
an accelerated detector depends on the number of spacetime dimensions 
offers an interesting feature that can possibly be utilized to detect
the extra dimensions\cite{stephens85}.
These very reasons also provide sufficient motivation to examine whether 
the results presented in this note are generic to other spacetimes which 
exhibit real or accelerated horizons (such as, for e.g., the black hole, 
de Sitter and the anti-de Sitter spacetimes)\cite{sriramip}.

\acknowledgements

The idea presented in this note originated during a discussion at 
the Second Workshop on Field Theoretic Aspects of Gravity held at 
the Radio Astronomy Centre, Ooty, India during October 3--9, 2001.
The author would like to thank the organizers for the invitation to 
speak at the Workshop.

\end{document}